# Continuously tunable dipolar exciton geometry for controlling bosonic quantum phase transitions

**Authors:** Zhenyu Sun[1], Haoteng Sun[1], Xiaohang Jia[1], An Li[1], Naiyuan J. Zhang[2], Ken Seungmin Hong[1], Joseph DePinho[3], Conor Y. Long[4], Kenji Watanabe[5], Takashi Taniguchi[6], Ou Chen[1], Jue Wang[7], Jia Li[8], Brenda Rubenstein[1,9], and Yusong Bai[1]*

**Affiliations:**

[1] Department of Chemistry, Brown University, Providence, RI, USA

[2] Department of Physics, University of Washington, Seattle, WA, USA

[3] School of Engineering, Brown University, Providence, RI, USA

[4] Department of Chemical and Biological Engineering, Princeton University, Princeton, NJ, USA

[5] Research Center for Electronic and Optical Materials, National Institute for Materials Science, Namiki, Tsukuba, Japan

[6] Research Center for Materials Nanoarchitectonics, National Institute for Materials Science, Namiki, Tsukuba, Japan

[7] Department of Physics, The Hong Kong University of Science and Technology, Hong Kong, China

[8] Department of Physics, University of Texas at Austin, TX, USA

[9] Department of Physics, Brown University, Providence, RI, USA

*Corresponding author. Email: yusong_bai@brown.edu

**ABSTRACT.** The geometry and binding energy of excitons, set by electron-hole wavefunction distributions, are fundamental factors that underpin their many-body interactions and determine optoelectronic properties of semiconductors. However, in typical solid-state systems, these quantities are fixed by material composition and structure. Here we introduce a polarizable interlayer exciton hosted in a two-dimensional tetralayer heterostructure whose dipole length, in-plane radius, and binding energy can be continuously programmed *in situ* over a wide range, enabling direct control over the nature of excitonic many-body phase transitions. An out-of-plane electric field redistributes layer-hybridized electron-hole wavefunctions, realizing *in situ* control of exciton geometry through a strong quadratic Stark response. This tunability further regulates the nature of interaction-driven Mott transition, transforming it from gradual to abrupt. Our results establish exciton geometry as a continuously tunable materials parameter, opening routes to exciton-based quantum phase-transition simulators and guiding the design of emergent optoelectronic functionalities from programmable excitonic materials.

Excitons, Coulomb-coupled electron-hole pairs, are elementary excitations in semiconductors that play a critical role in light-matter interactions and underpin a wide range of optoelectronic applications[1-3]. As neutral bosonic quasiparticles, they also offer an effective platform for exploring quantum many-body states[4-8]. In this context, exciton geometry, defined by the dipole moment and in-plane spatial extent of the electron-hole wavefunction, emerges as a fundamental parameter, as it governs binding energies,[2] optical transitions,[1] and the strength and nature of interparticle interactions.[6-8] Seminal examples include the crossover between Bose-Einstein condensation (BEC) and the Bardeen-Cooper-Schrieffer (BCS) state[6,9], which is induced by continuous variations of the Coulomb pairing strength and the size of the quasiparticles.

In recent years, optically addressable interlayer excitons (IXs) in two-dimensional (2D) semiconductor heterostructures have emerged as a promising bosonic testbed for exploring new excitonic phases and interaction-driven phase transitions, owing to their long lifetimes and enhanced interparticle interactions[4,5,10-12]. Because IXs consist of electron and hole wavefunctions largely localized in separate layers, the van der Waals stacking order and layer number introduce new degrees of freedom for tailoring exciton configurations. Examples range from dipolar excitons in twisted black phosphorus homostructures[13], where the dipole orientation can be selected via light polarization, to quadrupolar excitons assisted by every-other-layer quantum tunnelling[14-19]. While these studies have explored discrete tunability of exciton configurations, the continuous, *in situ* control over exciton dipole moment and spatial extent has remained experimentally inaccessible in semiconducting 2D materials. To date, such control has only been achieved in graphene double-layer systems under strong magnetic fields,[6,20] where Landau orbital wavefunctions of Coulomb-coupled electron-hole pairs can be adjusted, a mechanism not readily accessible in 2D semiconductors hosting optically addressable excitons.

Here we introduce a continuously tunable dipolar exciton with broadly adjustable sizes and binding energies formed in a tetralayer transition-metal dichalcogenide (TMD) heterostructure comprising a pair of valley-aligned homobilayers. In this system, an external electric field gradually redistributes the layer-hybridized electron and hole wavefunctions, simultaneously modulating the IX dipole length, in-plane radius, and binding energy *in situ*. This geometric tunability, manifested by a pronounced quadratic Stark response, yields large polarizability, allowing the IX static dipole moment to be varied nearly threefold. By tuning the IX geometry at fixed density, we access a previously unexplored regime in which the nature of inter-exciton interactions becomes continuously adjustable. As the dipole elongates and the in-plane radius expands, we observe a tunable quantum phase transition: the exciton Mott transition evolves from gradual to abrupt. In this way, we further address the long-standing puzzle regarding the intrinsic nature of density-driven exciton Mott transition. Together, these results establish exciton geometry, and its field-driven tunability, as a continuous control knob for bosonic many-body phase transitions, thereby opening routes to quantum-phase simulators based on dipolar excitons.

**Nonlinear Stark shift and highly polarizable interlayer excitons**

We fabricate dual-gate, type-II-aligned TMD heterostructures consisting of an R-stacked bilayer $WSe_2$ (R2L-$WSe_2$) on top of an R-stacked bilayer $WS_2$ (R2L-$WS_2$) (**Fig. 1a**). All adjacent layers in this tetralayer structure (R2L-$WSe_2$/R2L-$WS_2$) are aligned in the R-stacking configuration with small twist angles (~ 1 to 2°). We employ few-layer graphite to construct the dual-gate structure, where an out-of-plane electric field ($\vec{E_z} = E_z \hat{z}$) can be applied (see methods for fabrication details). In R2L-$WSe_2$/R2L-$WS_2$, we expect to observe optically generated IXs exhibiting polarizability, i.e., continuously tunable IX dipoles under an $E_z$ (**Fig. 1b**). As we establish below, the tunability arises from layer-hybridized electron and hole wavefunctions residing at the $WS_2$ conduction-band minimum (CBM) and $WSe_2$ valence-band maximum (VBM), which can be polarized by an electric field. This field-driven wavefunction redistribution modifies both the dipole length and in-plane extent of the IX, yielding a continuously tunable exciton geometry and binding energy (*vide infra*). This controllability provides a powerful means to manipulate excitonic many-body interactions.

The quadratic Stark shift of the exciton energy under $E_z$ directly reflects the continuously varying exciton dipole moment, i.e., its polarizability[21,22]. The IX PL in a conventional $WSe_2$/$WS_2$ hetero-bilayer exhibits a common linear Stark shift (**Fig. 1c**), consistent with a rigid dipole moment. In contrast, we observe a pronounced nonlinear Stark shift in the $E_z$-dependent IX PL spectra of the tetralayer heterostructure, indicating the presence of exciton polarizability (**Fig. 1d**). We emphasize that, this observed nonlinear Stark shift consistently evince across the overall sample area (**Fig. S1**), highlighting the homogeneity of the tetralayer heterostructure sample.

We now quantify the characteristic $E_z$-tuning of IX dipole moment in the heterostructures. For the hetero-bilayer, the data fits the linear Stark formula $\Delta E_{IX} = -e\vec{d} \cdot \vec{E_z}$ (**Fig. 1c**, gray solid line), where $\Delta E_{IX}$ is the peak energy shift, and $e\vec{d}$ denotes dipole moment. The fitting yields a dipole moment of ~ 0.61 $e$ nm (**Fig. 1c**, yellow solid line), consistent with the typical TMD hetero-bilayer spacing[11]. In the tetralayer device, however, a rigid dipole does not capture the full field dependence, and a piecewise model is required to fit the Stark shift. For $E_z$ > 28 mV nm$^{-1}$, the Stark shift again fits a linear model with a dipole moment of 0.57 $e$ nm (**Fig. 1d**, gray and yellow solid lines), suggesting an electron-hole pair localized at the central $WSe_2$/$WS_2$ interface. For -120 mV nm$^{-1}$ < $E_z$ < 28 mV nm$^{-1}$, the data is described by a second-order Stark formula: $\Delta E_{IX} = -e\vec{d_0} \cdot \vec{E_z} - \frac{1}{2}\alpha E_z^2$, where $\alpha$ defines the IX polarizability and $e\vec{d_0}$ denotes the dipole moment at zero electric field. From this fit (**Fig. 1d**, gray dash line; also see SI text for details), we extract an exciton polarizability of ~ 6.8 eV nm$^2$ V$^{-2}$, among the highest value observed for 2D dipolar excitons[22,23]. This large polarizability enables continuous tuning of the IX dipole moment over a wide range (**Fig. 1d**, yellow dash line). At $E_z = -120$ mV nm$^{-1}$, the dipole moment of 1.54 $e$ nm is the largest value reported in TMD heterostructures[12,24-28]. The effective dipole can be parked anywhere from 0.57 to 1.54 $e$ nm via an applied $\vec{E_z}$, distinguishing this R-stacked tetralayer system as a highly tunable low-dimensional platform for probing exciton-exciton interactions.

**Origins of IX polarizability**

The large IX polarizability in the R-stacked tetralayer system originates from layer-hybridized electron and hole wavefunctions in the homo-bilayer WS$_2$ and WSe$_2$, respectively. **Fig. 2a, b** show the $\vec{E_z}$-dependent reflectance-contrast (RC) spectra of R2L-WSe$_2$/R2L-WS$_2$, focusing on the WSe$_2$ (**Fig. 2a**) and WS$_2$ (**Fig. 2b**) A-exciton (X$_A$) spectral windows. In both cases, two oppositely shifting, linearly dispersing RC features are observed, crossing at $E_z = 0$ V nm$^{-1}$. Fitting these dispersions yields dipole moments of ~0.60 $e$ nm (**Fig. S4**), confirming their assignment as interlayer exciton transitions in the homo-bilayers (IX$_{Homo}$). Crucially, these field-dependent IX$_{Homo}$ features intersect the intralayer X$_A$ resonances and produce avoided crossings, indicating hybridization between interlayer and intralayer excitons (schematic in **Fig. S3**). This behavior suggests X$_A$ and IX$_{Homo}$ mix via spatially overlapping wavefunctions arising from layer-hybridized conduction- and valence-band states[29,30]. By fitting the avoided crossing spectral features, the hybridization energies are extracted to be ~11 meV for R2L-WSe$_2$ and ~8 meV for R2L-WS$_2$ (see SI). Notably, the absence of avoided crossings in the R-stacked WSe$_2$/WS$_2$ hetero-bilayer (**Fig. S5**) confirms that the observed hybridizations in R2L-WSe$_2$/R2L-WS$_2$ originates within each R-stacked homo-bilayer. Consistently, the extracted zero-field IX dipole moment of ~0.75 $e$ nm (**Fig. 1d**) in R2L-WSe$_2$/R2L-WS$_2$ exceeds the bilayer spacing (~0.60 $e$ nm), indicating intrinsic electron-hole delocalization and an extended exciton dipole even without an applied field.

Layer hybridization in the R-stacked homo-bilayers is enabled by spin-conserved band alignment at the K/-K valleys. Band structure calculations show that both the CBM and VBM reside at the K/-K points in R2L-WSe$_2$ and R2L-WS$_2$ (**Fig. S7**). Corresponding DFT calculations (**Fig. 2c**) reveal pronounced layer hybridization at both band edges. The wavefunction layer partitions indicate that the hybridization is stronger in R2L-WSe$_2$ than in R2L-WS$_2$ as shown in **Fig. 2c** (left two columns), which is consistent with the hybridization energies extracted from the avoided crossings in **Fig. 2a, b**.

As a control, we study a tetralayer composed of H-stacked natural bilayers (H2L-WSe$_2$/H2L-WS$_2$), where opposite valley spin orientations at K/-K prohibit interlayer hybridization in the homobilayers[31]. DFT calculations confirm that both the CBM and VBM wavefunctions remain layer-localized (**Fig. 2c**, right two columns). Consistently, the $E_z$-dependent RC spectra show no avoided crossings (**Fig. S6**), indicating the absence of layer-hybridized electronic states. Furthermore, the $E_z$-dependent PL spectra exhibit piecewise-linear Stark shifts (**Fig. 2d**), corroborating electric-field-driven switching between discrete dipolar exciton configurations rather than continuous dipole tuning. As shown in **Fig. 2d**, for $E_z > -61$ mV nm$^{-1}$, the Stark shift is captured by a fixed dipole moment of 0.60 $e$ nm, while for $E_z < -61$ mV nm$^{-1}$ a second linear regime emerges with a dipole moment of ~1.2 $e$ nm, comparable to the TMD trilayer spacing (see SI for detailed analysis). The abrupt transition between these two regimes highlights the absence of continuously tunable dipole moments in the H-stacked control and directly contrasts with the smooth dipole tunability enabled by valley-spin-allowed hybridization in R2L-WSe$_2$/R2L-WS$_2$.

**Electric-field control of IX radius and geometry-driven gradual-to-abrupt Mott crossover**

In 2D systems, the exciton Mott transition provides an effective means to estimate the exciton in-plane radius and probe the nature of exciton-exciton interactions[11,32,33]. We extract exciton in-plane Bohr radius using the 2D Mott criterion $1/\pi \sim n_{Mott} a_B$[11,34], which relates the critical exciton ionization density $n_{Mott}$ to exciton Bohr radius $a_B$. A key spectroscopic signature of the exciton Mott transition is the excitation-density ($n_{eh}$) dependence of the PL linewidth broadening[11,32,33,35]. We therefore examine $n_{eh}$-dependent PL spectra of IXs while tuning their dipole moment with an out-of-plane field $E_z$.

We thoroughly examine CW $n_{eh}$-dependent peak-normalized PL spectra of the R2L-WSe$_2$/R2L-WS$_2$ heterostructure over $E_z$ ranging from 45 mV nm$^{-1}$ to -85 mV nm$^{-1}$, corresponding to IX dipole moments varying from 0.57 to 1.32 $e$nm (**Fig. S10, S11**). **Fig. 3a-c** display such results at three representative fields ($E_z$ = 30, 0, and -53 mV nm$^{-1}$). The $n_{eh}$ spans 1.2×10$^{10}$ cm$^{-2}$ to 6.8×10$^{12}$ cm$^{-2}$. Densities are calibrated by relating CW and pulsed excitation conditions (see SI Text and **Fig. S9**) following established protocols[5]. **Fig. 3d** shows the constructed full-width-at-half-maximum (FWHM) map across $E_z$ and $n_{eh}$: at the lowest $n_{eh}$, the IX linewidth remains uniformly narrow across all $E_z$, with a FWHM of ~10 meV, whereas the linewidth increases by a factor of about four to six with increasing $n_{eh}$ (depending on $E_z$), verifying exciton ionization[11,33]. We emphasize that, unlike bilayer systems, the R2L-WSe$_2$/R2L-WS$_2$ heterostructures exhibit a flat-land morphology with no discernible moiré features, as evidenced by the PFM image and the smooth, doping-dependent evolution of the RC spectra (**Fig. S12**). This unique feature makes the system an ideal platform for investigating IX-IX interactions without complications from moiré potential traps.

The exciton in-plane radius grows with increasing IX out-of-plane dipole moment. The density-driven linewidth evolution in **Fig. 3d** is quantitatively captured by an S-shaped logistic function (see SI Text and **Fig. S14, S15**), which yields the Mott density $n_{Mott}$ and the transition width $\Delta n_{eh}$ (**Fig. 3e** and **Table S1**). As summarized in **Fig. 3e**, $n_{Mott}$ progressively shifts toward the lower excitation-density regime as the electric field is swept from positive to negative. The progressive field-driven decrease of $n_{Mott}$ allows us to determine an expansion of the IX in-plane radius $a_B$ using the 2D Mott criterion $1/\pi \sim n_{Mott} a_B$. Additionally, the corresponding IX binding energy $E_b$ can be estimated from the relation $E_b \propto \hbar^2/2M_r a_B^2$[8], where $M_r$ is the reduced mass. **Fig. 3f** shows variations of $a_B$ and $E_b$ as functions of the out-of-plane field $E_z$ and the static IX dipole moment $e\vec{d}$: increasing $|E_z|$ continuously enlarges $|e\vec{d}|$ and expands the in-plane $a_B$ from ~3 nm to ~12 nm, accompanied by a monotonic decrease of the effective binding energy to ~6 % of that of a regular bilayer IX. The observed modulations of $E_b$ and $a_B$ upon tuning the IX dipole moment are highly consistent with previous theoretical predictions[8]. Additionally, steady-state pump-probe spectroscopy reveals a pronounced plasma-gain feature[11,36] at densities closely matching $n_{Mott}$ extracted from PL linewidth broadening (SI Text and **Fig. S18**). This correspondence confirms that the explored density regime spans the Mott transition and validates PL linewidth as a reliable signature of exciton ionization.

We observe an IX-geometry-driven gradual-to-abrupt Mott crossover, underscoring that exciton geometry governs the nature of exciton-exciton interactions. As highlighted in **Fig. 3e** (and **Table S1**), the transition width $\Delta n_{Mott}$ (the density span over which Mott ionization occurs), obtained from logistic fitting (SI Text and **Fig. S14, S15**), systematically narrows with increasing dipole moment and radius. This affirms that large IXs undergo a more abrupt interaction-driven Mott ionization and demonstrates a geometry-driven gradual-to-abrupt Mott crossover that is continuously tunable by the applied field. Accordingly, the $\partial$FWHM/ $\partial n_{eh}$ map (**Fig. S16**) reveals a clear boundary for $E_z <$ -40 mV nm$^{-1}$. Likewise, the 30, 40, and 50 meV linewidth contours in **Fig. 3d** converge as $E_z$ is tuned negative, revealing that the density required to reach a given linewidth decreases as the IX geometry expands. At the largest dipole moment and radius ($|e\vec{d}| \approx 1.30\ e$ nm, $a_B \approx 12$ nm), these contours begin merging, indicating an abrupt density-driven linewidth increase.

Notably, our observations also address a long-standing question concerning the intrinsic nature of the density-driven exciton Mott transition. Experimentally, this transition has been reported to evolve either gradually with increasing density[35,37] or to occur abruptly as an exciton collapse[32,38], leaving unresolved whether the transition is intrinsically gradual or abrupt. The observed IX geometry-driven gradual-to-abrupt Mott crossover reconciles these prior discrepancies and identifies exciton geometry and binding energy as relevant parameters affecting the nature of this many-body process. The gradual-to-abrupt evolution is not observed in the $E_z$-$n_{eh}$-FWHM map of WSe$_2$/WS$_2$ hetero-bilayer (**Fig. S13**), highlighting the essential role of field-programmable IX geometry and binding energy in shaping IX-IX interactions.

Complementary evidence for the essential role of IX geometry in steering IX-IX interactions is the $E_z$-driven Mott transition. At low density (~4.5×10$^{10}$ cm$^{-2}$), the PL linewidth remains narrow and nearly constant across the examined field range (**Fig. 3g** and **Fig. S19a**), indicating a bound IX regime; the $E_z$-dependent PL peak energy also follows the IX nonlinear Stark shift (**Fig. 1d**). At intermediate density (~1.5×10$^{12}$ cm$^{-2}$), however, driving $E_z$ negative induces linewidth broadening: the IX peak evolves into a ~70 meV-wide featureless band (**Fig. 3g** and **Fig. S19b**), characteristic of plasma-like recombination from continuum. Concurrently, the PL peak deviates from the nonlinear Stark trajectory and bends to higher energy, indicating field-driven exciton ionization. As the IX geometry is modulated with more negative $E_z$, a growing fraction of excitons dissociates into free carriers. Once ionized, these carriers screen the external field, causing the Stark nonlinearity to collapse; the system behaves as a charged capacitor rather than a gas of dipoles, and well-defined excitonic dipole moments no longer exist. Importantly, because $n_{eh}$ is calibrated to remain constant across the $E_z$ sweep (**Fig. S9**), this transition is purely field-driven. Together, these observations demonstrate that the excitonic phase can be tuned not only by density but also by field-programmed IX geometry and binding energy.

**Microscopic view of tunable exciton-exciton interactions**

We now discuss the microscopic origins of the tunable IX-IX interactions evinced in the Mott transition. The distinct geometry of IXs and the contrasting nature of IX-IX interactions are schematically illustrated in **Fig. 4a**. The continuous reduction of IX binding energy plays a key role in regulating the evolution from a gradual to an abrupt Mott transition. When IXs are strongly bound, the large $E_b$ provides an energetic buffer that stabilizes the excitonic state: as density increases, Coulomb screening continuously weakens $E_b$, but the large initial $E_b$ ensures that ionization proceeds continuously along the $n_{eh}$ axis. By contrast, when IXs are weakly bound, the bound state lies close to the continuum, and even a modest enhancement of screening can collapse $E_b$. Once a few excitons ionize, the resulting free carriers enhance screening further, accelerating ionization and establishing a runaway feedback loop[33,34]. The smaller the $E_b$, the lower the barrier protecting the bound state and the easier it becomes for this feedback to trigger a collective, avalanche-like transition. Consistently, in the R2L-WSe$_2$/R2L-WS$_2$ heterostructure, the PL evolution transforms from a slow, gradual shift at small dipole moments to an abrupt ionization at large dipole moments. This observation aligns with previous theoretical suggestions that reducing $E_b$ promotes the screening feedback that converts a gradual crossover into an abrupt Mott-like transition[39].

The increased IX radius itself can provide an additional key mechanism contributing to the observed abruptness through exchange repulsion. Based on the $E_z - n_{eh} -$ FWHM map in **Fig. 3d**, we identify a dilute excitation-density range $n_{eh} < 2 \times 10^{11}$ cm$^{-2}$, where the excitonic phase dominates. We then map the optical stiffness, $\partial E_{PL}/\partial n_{eh}$ (i.e., the optical energy penalty for adding IXs to the system), as a function of $E_z$ and $n_{eh}$ (**Fig. 4b**). Importantly, in this dilute regime, large-size IXs (e.g., $E_z < -50$ mV nm$^{-1}$) exhibit stiffness roughly an order of magnitude higher than small-size IXs ($E_z > 0$ mV nm$^{-1}$) (also see **Fig. S20**). Such a contrasting optical stiffness cannot be explained by classical dipole-dipole repulsion alone ($\propto 4\pi n_{eh} e^2 d_{IX}/\epsilon$)[40], since the dipole length differs by only a factor of two between these field regimes (see SI Text). Moreover, the consistently narrow PL linewidths and well-defined nonlinear Stark trajectories (**Figs. 3g** and **S19**) in this density range exclude ionization and band-gap-renormalization effects as possible sources of the large $\partial E_{PL}/\partial n_{eh}$. Given the pronounced in-plane Bohr-radius extension under negative fields (**Fig. 3f**), the excess stiffness likely reflects a wavefunction-overlap-driven interaction that strengthens with exciton size. As the in-plane radius expands, the overlap between electron (and hole) envelopes belonging to different excitons increases during dynamic scattering, enhancing Pauli exchange repulsion[41,42]. Upon ionization into the plasma phase, the system can relieve this exchange penalty associated with the large IX radius by forming a free-carrier fluid with weaker interactions. We emphasize that the role of IX wavefunction delocalization and enhanced exchange repulsion deserves further theoretical investigation to elucidate its quantitative impact on the abruptness of the interaction-driven Mott transition.

**Fig. 4c** places our electrically tunable IX geometry in the context of previously demonstrated platforms. Previous studies in TMD bilayers, TMD multilayers, and double-graphene heterostructures have achieved either modest variations in exciton dipole moment or access to excitons with relatively large in-plane radii, but the two degrees of freedom have remained largely decoupled and limited in range. The tetralayer system here enables a continuous and simultaneous tuning of both the IX dipole moment and the in-plane radius over a substantially broader parameter space. The accessible $|e\vec{d}|$ values span nearly a factor of three, while the Bohr radius $a_B$ expands fourfold, reaching ~12 nm. This combined tunability of vertical dipole length and lateral spatial extent establishes a new regime of exciton engineering, providing a programmable geometry that directly enables the control of many-body phenomena.

**Conclusion and outlook**

This work establishes a new interlayer-exciton platform whose geometry (i.e., dipole moment and in-plane radius) and binding energy can be continuously programmed by an external electric field. By reshaping the excitonic wavefunction *in situ*, we gain direct control over exciton-exciton interactions, accessing regimes that are fundamentally inaccessible in systems with fixed exciton properties. Using the exciton Mott transition as a diagnostic, we show that expanding exciton geometry qualitatively alters the many-body behavior, transforming the transition from gradual to abrupt. More broadly, electric-field programmability of IX geometry enables continuous traversal of the exciton phase diagram along density and geometry axes, providing a direct route to exploring correlated excitonic phases, including dipolar crystals, supersolids, and quantum liquids, within a single platform, as predicted by theories[7,8,43,44]. Our results establish exciton geometry as a continuously tunable materials parameter, on par with density and temperature, enabling programmable excitonic materials and systematic exploration of interaction-driven quantum phase transitions in 2D bosonic systems.

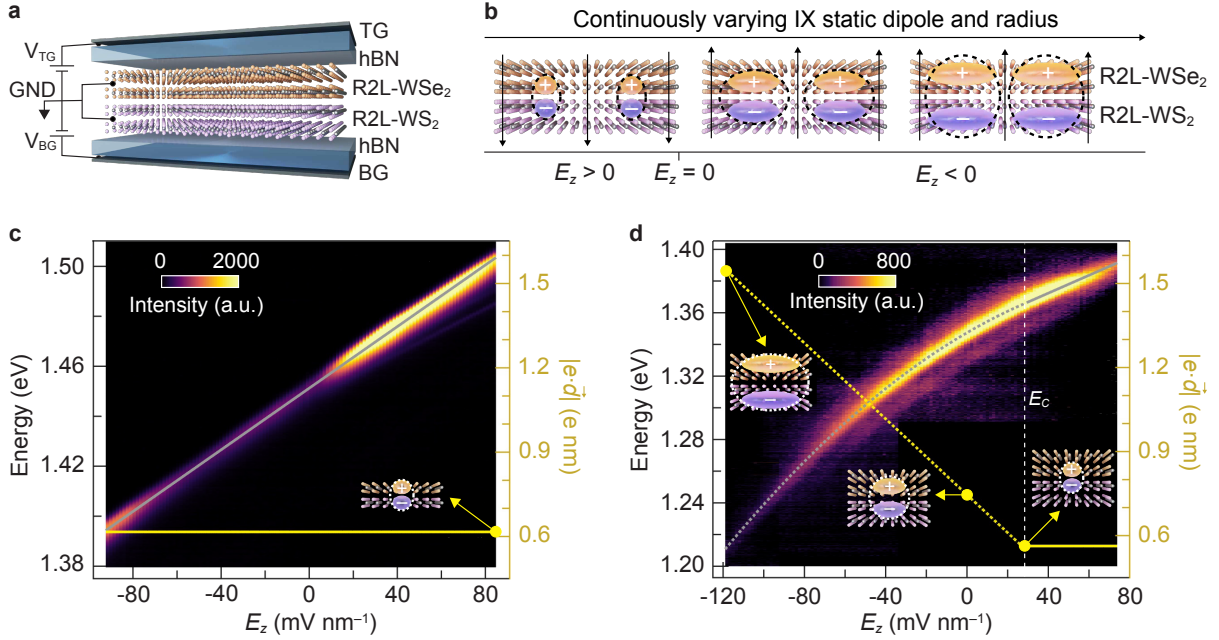

**Fig. 1 | Interlayer exciton polarizability in a 2D hetero-tetralayer system. (a)** Schematic of a dual-gate R2L-WSe$_2$/R2L-WS$_2$ device. Both top and bottom gates consist of few-layer graphite (3-5 layers); the gold contact to the TMD is bridged by a graphite flake. **(b)** Illustration of the continuously tunable interlayer exciton (IX) dipole moment under an out-of-plane electric field $E_z$ (gray arrows). When $E_z$ reaches a critical positive value (28 mV nm$^{-1}$), electrons and holes localize in the middle layers, forming a bilayer-limit dipolar IX. **(c, d)** $E_z$-dependent PL spectra of IXs in WSe$_2$/WS$_2$ **(c)** and in the hetero-tetralayer **(d)**, measured at charge neutrality (Methods). Gray solid lines in **(c)** and **(d)** are linear Stark-shift fits; gray dashed curve in **(d)** is quadratic Stark-shift fit. The vertical white dashed line (∼ 28 mV nm$^{-1}$) in **(d)** marks the field beyond which the tetralayer IX dipole moment reaches the bilayer limit, as schematically illustrated on either side. Yellow solid and dashed lines in **(c)** and **(d)** represent the first-order derivative of the $E_z$-dependent PL peak energy; particularly, yellow dashed line in **(d)** highlights the continuously tunable IX dipole moment ($e\vec{d}$) in the hetero-tetralayer structure. All spectra were acquired at 2 K with 500 nW continuous-wave (CW) excitation ($h\nu$ = 2.33 eV).

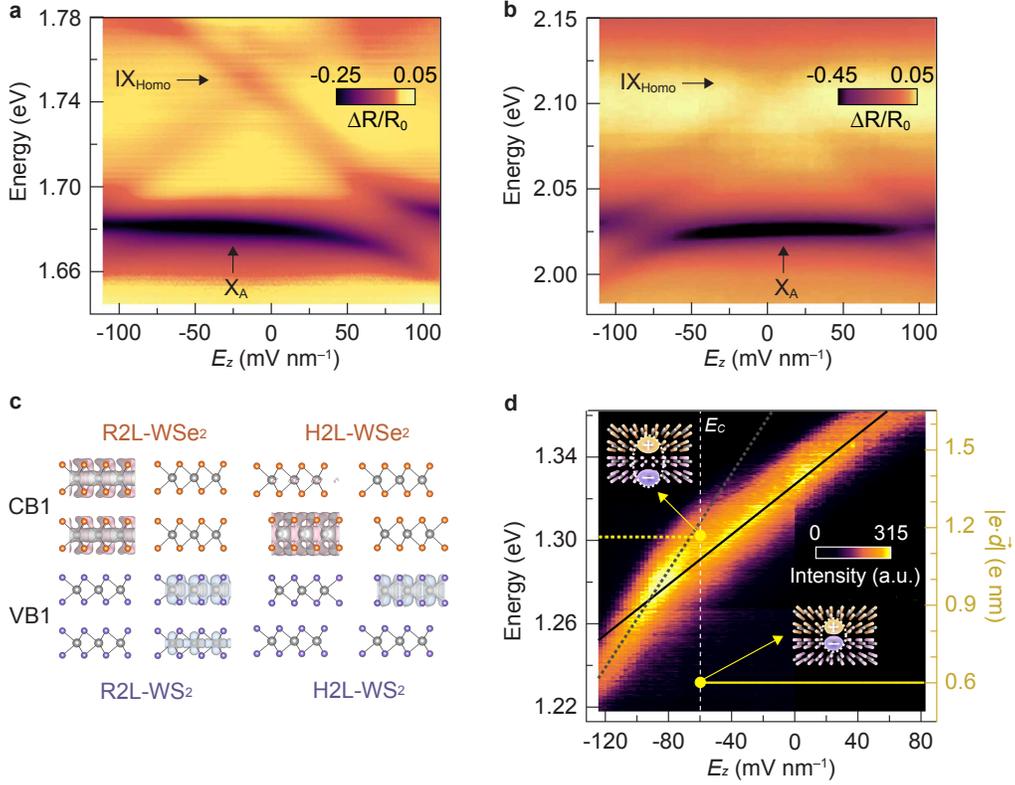

**Fig. 2 | Interlayer hybridization in the hetero-tetralayer system. (a, b)** Reflectance contrast (RC) spectra versus out-of-plane electric field $E_z$ for the WSe$_2$ (**a**) and WS$_2$ (**b**) A-exciton regimes in the R2L-WSe$_2$/R2L-WS$_2$ hetero-tetralayer sample, measured at 2 K. **(c)** Spatial distributions of the electronic wavefunctions at the *K*-valley valence and conduction band edges for the R2L-WSe$_2$/R2L-WS$_2$ (left) and H2L-WSe$_2$/H2L-WS$_2$ (right) stacks. The corresponding DFT-calculated band dispersions are shown in **Fig. S7**. **(d)** $E_z$-dependent PL spectra of IXs in H2L-WSe$_2$/H2L-WS$_2$, acquired with 1 µW CW excitation at 2 K ($h\nu$ = 2.33 eV). Solid and dashed black lines in (**d**) are linear Stark shift fits, highlighting two distinct responses corresponding to IX dipole moments of 1.16 *e* nm (yellow dashed line) and 0.60 *e* nm (yellow solid line).

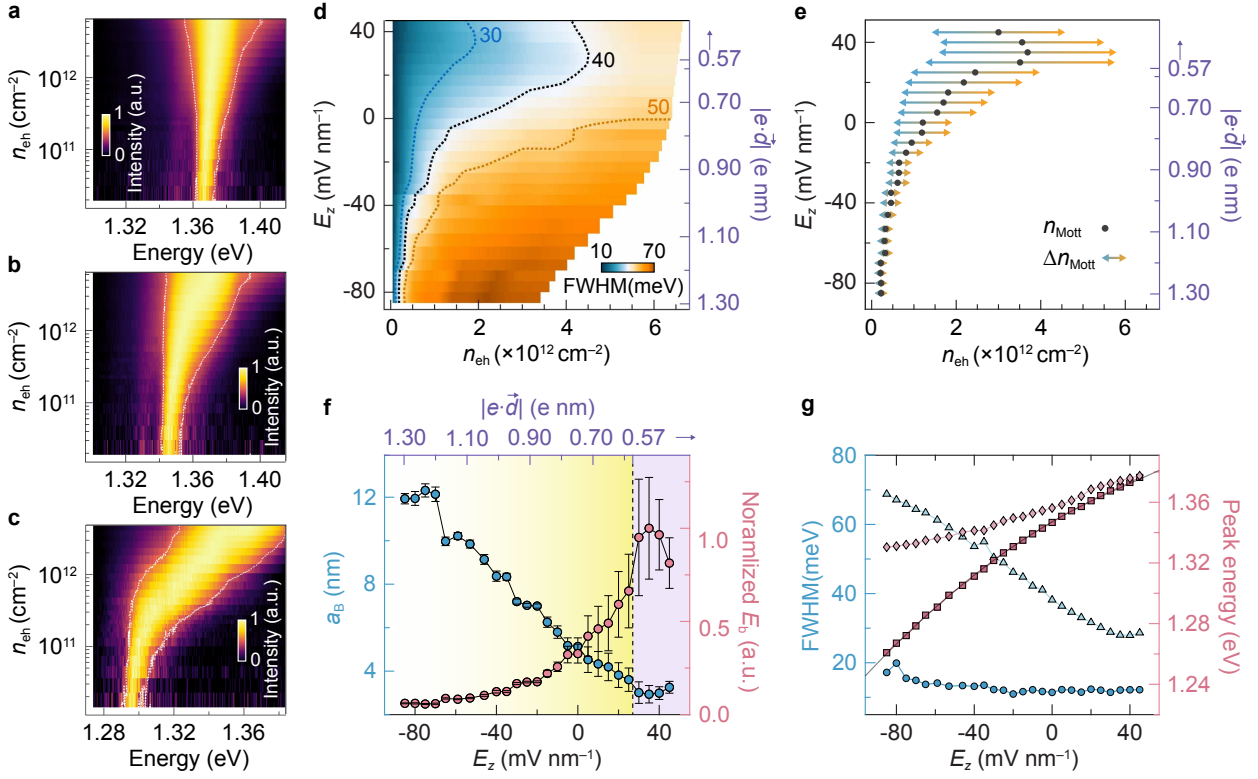

**Fig. 3 | Field control of IX radius and geometry-driven gradual-to-abrupt Mott crossover. (a-c)** Excitation-density ($n_{eh}$)-dependent PL spectra of IXs in the R2L-WSe$_2$/R2L-WS$_2$ sample, acquired under out-of-plane electric fields $E_z = 30$ (**a**), 0 (**b**), and −53 (**c**) mV nm$^{-1}$. The excitation density spans $1.2 \times 10^{10}$ cm$^{-2}$ to $6.8 \times 10^{12}$ cm$^{-2}$. **(d)** FWHM map plotted versus $E_z$ and $n_{eh}$ for the same sample. The right axis shows the corresponding dipole moment $|e\vec{d}|$; for $E_z > 28$ mV nm$^{-1}$, $|e\vec{d}|$ saturates at 0.57 $e$ nm (purple arrow). Dashed contours mark constant FWHM values of 30 (blue), 40 (black), and 50 (orange) meV. **(e)** Mott transition density range $\Delta n_{eh}$ (blue-orange double arrows) at different electric field $E_z$. The black solid circles mark $n_{Mott}$. $\Delta n_{eh}$ is obtained from logistic fitting of the $n_{eh}$-dependent FWHM (SI Text). **(f)** Estimated in-plane Bohr radius ($a_B$, purple) and binding energy ($E_b$, orange) of IXs in R2L-WSe$_2$/R2L-WS$_2$ as functions of out-of-plane electric field $E_z$ (bottom axis) and corresponding dipole moment $|e\vec{d}|$ (top axis). Error bars denote uncertainty from the fitted Mott density. The black dashed line marks the field beyond which the tetralayer IX dipole moment reaches the bilayer limit. **(g)** $E_z$-dependent PL FWHM (blue symbols) and peak energy (red symbols) measured at fixed excitation densities. Dark blue circles and dark red squares correspond to $n_{eh} = 4.5 \times 10^{10}$ cm$^{-2}$; light blue triangles and light red diamonds correspond to $1.5 \times 10^{12}$ cm$^{-2}$; the black solid curve represents the Stark-shift trajectory extracted from **Fig. 1d**.

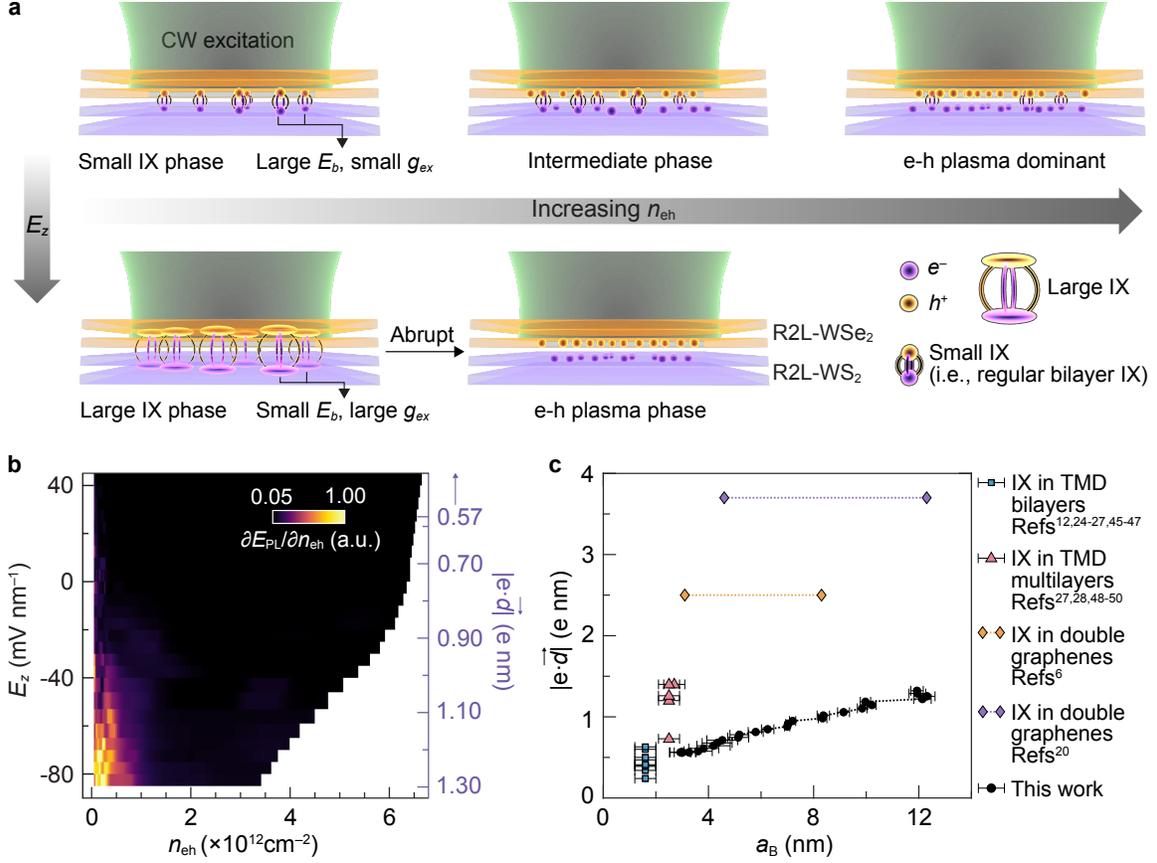

**Fig. 4 | Excitonic interactions and the tunable IX geometry. (a)** Schematic illustration of the electric-field-induced modulation of the IX geometry in the tetralayer and the resultant varying nature of IX-IX interactions. Increasing $E_z$ redistributes the layer-hybridized electron and hole wavefunctions, enhancing the IX dipole moment $|e\vec{d}|$ and expanding the in-plane Bohr radius $a_B$. Tightly bound, small-size IXs display a softened interaction strength with increasing density, whereas the weakly bound large-size IXs exhibit an abrupt response to increasing density. **(b)** Optical stiffness map ($\partial E_{PL}/\partial n_{eh}$) plotted versus $E_z$ and $n_{eh}$, where $E_{PL}$ is the PL peak energy. The right-hand axis in **(b)** denotes IX dipole moment. Increasing $E_z$ increases both the IX size and the optical stiffness, revealing possible cooperative tuning of exciton binding and exchange-driven IX-IX interactions. **(c)** Comparison of IX dipole moment and in-plane radius across previously reported IX systems (TMD bilayers, multilayers, and double-graphene heterostructures) and the present hetero-tetralayer platform. Blue squares denote IXs formed between adjacent layers[12,24-27,45-47]. Red triangles represent IXs extending across multiple layers[27,28,48-50]. $a_B$ is extracted following the model in Ref[8]. The orange and purple diamonds label IXs in double-graphene structures separated by thin hBN spacers[6,20]; horizontal dotted lines indicate the experimentally tunable $a_B$ range via magnetic field. Black circles show our R2L-WSe$_2$/R2L-WS$_2$ results: the hetero-tetralayer platform uniquely provides simultaneous, continuous, and broad-range tunability of both $|e\vec{d}|$ and $a_B$.

## Methods

### Device fabrication

All 2D heterostructure samples were fabricated using a layer-by-layer van der Waals stacking technique. Monolayers of WSe$_2$ and WS$_2$ (HQ Graphene) were mechanically exfoliated from bulk crystals. Hexagonal boron nitride (hBN) flakes with thicknesses of 20-50 nm and atomically flat surfaces were also obtained by mechanical exfoliations. All flakes (WSe$_2$, WS$_2$, and hBN) were characterized by AFM and/or Raman spectroscopy prior to their use in device fabrications. A general stacking protocol is described in the following. A transparent polydimethylsiloxane (PDMS) stamp coated with a thin layer of polycarbonate (PC) was employed to sequentially pick up the following layers: top hBN flake, few-layer graphene top gate, top hBN encapsulation flake, graphite contact, TMD layers, bottom hBN encapsulation flake, few-layer graphene bottom gate, and bottom hBN flake. For preparing the R2L-WSe$_2$/R2L-WS$_2$ heterostructure, relatively large-area WSe$_2$ and WS$_2$ parental flakes were selected and then divided into two separate pieces by AFM cutting. Then, the resulting R2L-WSe$_2$ and R2L-WS$_2$ pieces could be subsequently stacked with high twist-angle accuracy. The completed heterostructure was transferred onto a clean silicon wafer with a 285-nm SiO$_2$ layer to provide optimal optical contrast, using an elevated transfer temperature of ~170-180 °C. Residual PC was removed with successive rinses of chloroform, acetone, and isopropanol. Metal-electrode patterns for connecting the graphite gates and contacts were defined by electron-beam lithography (FEI Helios 600 NanoLab) and reactive ion etching (PlasmaTherm Vision 320), followed by metal deposition of 5-nm Cr/100-nm Au (Angstrom electron-beam evaporation system).

### Optical measurements

All optical measurements were performed using a home-built confocal microscope system (**Fig. S17**) integrated with a helium-recirculating optical cryostat (Quantum Design OptiCool, base temperature ~2 K) and a 100×, NA = 0.75 objective (Zeiss LD EC Epiplan-Neofluar 100×/0.75 DIC M27). Unless otherwise specified, device samples were maintained at ~2 K under vacuum during all experiments. A XYZ nanopositioner (Attocube) was used to position the sample, and a scanning Galvo system (Thorlabs GVS012) within the confocal setup enabled precise control over the excitation and signal-collection area. Continuous-wave (CW) excitation was provided by a 532 nm laser diode (Thorlabs DJ532-40). For time-resolved measurements, excitation was achieved with a 730 nm pulsed laser (Light Conversion ORPHEUS-N-2H, pumped by PHAROS PH2-10W, 250 kHz). A stabilized tungsten-halogen lamp (Thorlabs SLS201L) served as the broadband white-light source for reflectance spectroscopy. With the cryostat objective, laser beams were focused to a ~1 μm spot size, while lamp light was collimated and reshaped to achieve a diffraction-limited spot size (~1 μm). Steady-state PL and reflectance signals were collected and spectrally resolved using a spectrograph (Princeton Instruments HRS-500) equipped with a 600 g/mm diffraction grating and detected with a cryogenically cooled InGaAs array camera (Princeton Instruments PyLoN-IR). Time-resolved PL was measured by time-correlated single-photon counting, using a single-photon counting module (Excelitas SPCM-AQRH) synchronized with the pulsed laser and recorded with a time-tagging electronics module (PicoQuant TimeHarp 260).

**Electric field control**

The equivalent circuit of our dual-gate device was a parallel plate capacitor with a 2D sheet of material (i.e., the TMDC heterostructure) in the middle. The dielectric constants for h-BN and TMDC are $\varepsilon_{h-BN} \approx 3.9$, and $\varepsilon_{TMDC} \approx 7.2$[12,15]. Then, the electric field generated in between the top and bottom gates can be expressed as $E_z = \frac{V_{tg} - V_{bg}}{d_{hBN}} \cdot \frac{\varepsilon_{h-BN}}{\varepsilon_{TMDC}}$, where $V_{tg}$ and $V_{bg}$ are the top and bottom gate voltages, respectively; $d_{hBN}$ is the total thickness of top and bottom hBN. Prior to scanning $E_z$, we performed a gate-gate reflectance map, to better identify the ratio of $V_{tg}/V_{bg}$ that ensures a pure $E_z$ without doping effect (**Fig. S2**). Gate voltages were supplied via two identical source meters (Keithley 2450, source accuracy ~ 0.012%, calibrated) which were computer controlled through a GPIB bus.

**DFT calculations**

Spin-orbit-coupled density functional theory (DFT+SOC) calculations were carried out using the plane-wave code Quantum Espresso[51-53]. We used the Perdew–Burke–Ernzerhof (PBE) functional of the generalized-gradient approximation (GGA)[54] as the exchange-correlation energy. Core–valence interactions were treated by full relativistic projector-augmented-wave (PAW) pseudopotentials[55]. A kinetic energy cutoff of 100 Ry was applied to the wavefunctions. The two-dimensional heterostructures were modeled with an in-plane lattice constant of $a = 3.19$ Å and a vacuum spacing of at least 20 Å along the out-of-plane direction to suppress interlayer interactions between image atoms. $14 \times 14 \times 1$ $k$-mesh is employed in the first Brillouin-zone integrations. Electronic self-consistent calculation was converged to $1 \times 10^{-10}$ Ry in total energy, and atomic positions were relaxed until the residual forces on all atoms were below $1 \times 10^{-3}$ Ry/Bohr. Long-range dispersion interactions were included via the DFT-D3 correction[56].

**Data availability**

The data that support the findings of this study are presented in the article and its Supplementary Information. Further data are available from the corresponding author upon request.

**Author contributions**

Y.B. conceived the research and oversaw the project. Z.S. fabricated the devices and performed the optical experiments with assistance from Y.B., H.S., A.L., N.J.Z., K.S.H., J.D., C.Y.L., and J.W. X.J. and B.R. performed the theoretical calculations. K.W. and T.T. grew the hBN crystals. Y.B. and Z.S. analyzed the data with inputs from O.C., J.W., and J.L. All authors discussed the results and commented on the manuscript.

**Competing interests:** The authors declare no competing interests.

**Data and materials availability:** All data are available in the main text or the supplementary materials.